\documentclass[12pt]{iopart}
\usepackage{iopams}  
\usepackage{graphicx}
\usepackage{dcolumn}
\usepackage{bm}
\usepackage{hyperref}

\def\v#1{{\bf#1}}
\def\be{\begin{equation}}
\def\ee{\end{equation}}
\def\bea{\begin{eqnarray}}
\def\eea{\end{eqnarray}}

\def\ncal{\mbox{$\cal N\,$}}

\def\wcal{\mbox{$\cal W\,$}}
\def\<{\langle}
\def\>{\rangle}

\begin{document}
\nocite{*}

\title{Spontaneous focusing as an emergent phenomenon}

\author{E. Sadurn\'i}
\address{Instituto de F\'isica, Benem\'erita Universidad Aut\'onoma de Puebla,
Apartado Postal J-48, 72570 Puebla, M\'exico}
\ead{sadurni@ifuap.buap.mx}

\begin{abstract}
We analyze the emergence of diffractive focusing in the transition from discrete to continuous space-time variables. Three types of dynamical equations are studied in a top-to-bottom approach, starting with the most general system. First we solve a linear cellular automaton of two species, then a nearest neighbour tight-binding array and finally the time-dependent Schr\"odinger equation. All models are shown to produce diffractive solutions for square packet distributions. The main result of this paper is that in discrete variables, the nature of the solutions depends strongly on the size of wavepackets, whereas in continuous variables, diffraction due to discontinuities exists at every scale. A transition in the number of participants or cells is identified by means of a measure and the corresponding phenomenon is further analyzed by a generalization of Wigner functions to discrete variables and crystals. 
\end{abstract}




\section{Introduction}

Diffractive phenomena are the result of special boundary conditions in wavefunctions and their derivatives with respect to space and time \cite{bruckner1997},\cite{bestle1995}. This is a quite general statement that involves various types of waves, such as vibrations \cite{larsen1981}, \cite{mow1971}, \cite{graff1991}, electromagnetic fields \cite{born1980} and quantum-mechanical probability amplitudes \cite{moshinsky1952}. One of the hallmarks of diffraction is an unusual increase of intensities around sharp edges and corners, accompanied by strong oscillations. With two or more of such discontinuities combined, a mechanism of focusing emerges \cite{case2012}.

\begin{figure}[h!]
\begin{center}  \includegraphics[width=8cm]{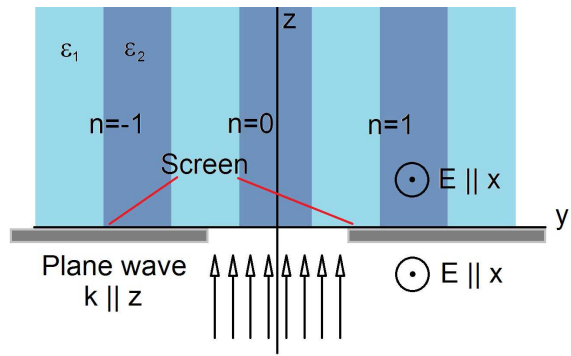} \end{center}
\caption{\label{fig:-1}An example of a tight-binding realization using a periodic array of dielectric layers with permittivities $\varepsilon_1$, $\varepsilon_2$. An electromagnetic wave impinging parallel to the slabs is partially blocked by absorptive screens. Inside the medium we have  a TE mode in $x$ direction. The wave  propagates along $z$ as a fictitious time and $y$ as space. The index $n$ is our effective discrete variable, and it labels the localized resonance in each layer $\varepsilon_2 > \varepsilon_1$. Since only one resonance is considered, only one band contributes to the dynamics.}
\end{figure}

The elements that allow diffraction, as quoted above, seem to be inherent to continuous space-time variables. It is natural to ask whether these wave-like effects appear in discrete systems, such as cellular automata \cite{molofsky1994}, \cite{volpert2009}, \cite{kaitala2000}, spatially discretized Schr\"odinger and Helmholtz equations \cite{sadurni2012(2)}, \cite{sadurni2013}, nearest-neighbour tight-binding models for matterwaves in lattices \cite{bloch2005}, \cite{oberthaler1996}, \cite{morsch2006} and even coupled optical waveguides in the regime of paraxiality \cite{chien2007}, \cite{russell2003}. A proposal involving dielectric layers is shown in fig. \ref{fig:-1}. While the transition from discrete to continuous can be ensured by reducing a step size or by increasing coupling constants between cells, the goal of this paper is to show that diffraction also emerges by increasing the number of {\it participants.\ } The size of intial wavepackets, counted in cell units, shall be the main control parameter of the present study. Interestingly, such a quantity corresponds to the number of initially lit up optical guides assembled in a chain, or the initial population of a prey-predator cellular automaton. 

We present our discussion in the following order: In section 2 we present diffractive solutions for square packets in discrete and continuous variables hierarchically, starting with a fully discretized automaton of two species in  position and time. Then we specialize the results to continuous time, leading to diffraction in tight-binding arrays and finally we touch upon the continuous limit leading to a Schr\"odinger equation. In section 3 we quantify the transition by means of an intuitive measure, corresponding to the probability that a quantum-mechanical particle enters into a small interval around the origin. In section 4 we make use of the Wigner function and its tight-binding generalization, with the aim of characterizing the motion in phase space or Bloch momenta and site indices for the discrete case. We comment on our results in section 5. 

\section{Three nested dynamical models}

We proceed to describe three dynamical models starting from a fully discretized, two-species linear cellular automaton (TSLA), then we specialize it to a nearest neighbour tight-binding model (NNTBM) in continuous time, and finally we recover a one-dimensional time-dependent Schr\"odinger equation (1DSE) or a two-dimensional Helmholtz equation in the paraxial approximation. The properties of such dynamical models are summarized in table \ref{tab:table1}.
\begin{table}[t]
\caption{\label{tab:table1}%
A classification of our discrete equations and their limits.
}
\begin{tabular}{lccc}
\br
\textrm{ -- }  &
\textrm{TSLA}&
\textrm{NNTBM}&
\textrm{1DSE} \\
\mr
\textrm{Time} & Discrete & Continuous & \textrm{Continuous} \\
\textrm{Space} &  Discrete &  Discrete & \textrm{Continuous} \\
\textrm{Evolution} & Non-unitary & Unitary & \textrm{Unitary}\\
\textrm{Parameters} & Time and space steps & Neighbour coupling & \textrm{Dimensionless}\\
\textrm{Diffractive if} & \begin{tabular}{c}Large number of cells\end{tabular}& Large wavepackets& \textrm{Always} \\
\end{tabular}
\end{table}

\begin{figure}[h!]
\begin{center}  \includegraphics[width=11.5cm]{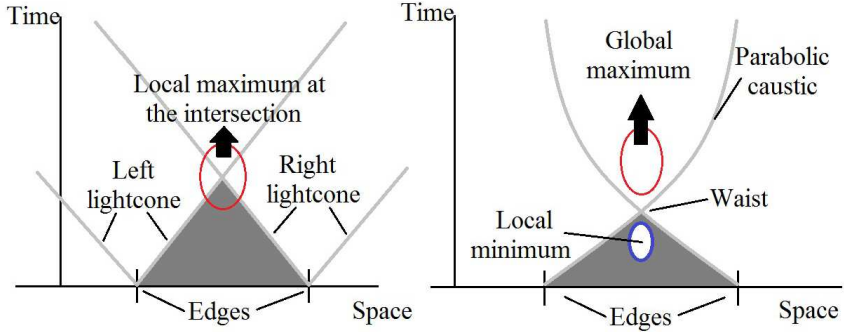} \end{center}
\caption{\label{fig:1.2}A comparison of space-time diagrams and their anatomy for non-focusing (left) and focusing (right) distributions.}
\end{figure}

\subsection{Emergent focusing in a linear automaton}

The TSLA is a type of cellular automaton that can be solved analytically. Let $x_{\tau,n}, y_{\tau,n}$ be real quantities describing populations of two species $x,y$ at discrete time $\tau$ and discrete position $n$. When the growth rate of one species depends on the other at the surroundings of a certain point, we have a discrete diffusion-reaction system; this type of model can be identified with prey-predator dynamics when diffusivities have opposite signs. We have

\begin{figure}[h!]
\begin{center}  \includegraphics[width=9.5cm]{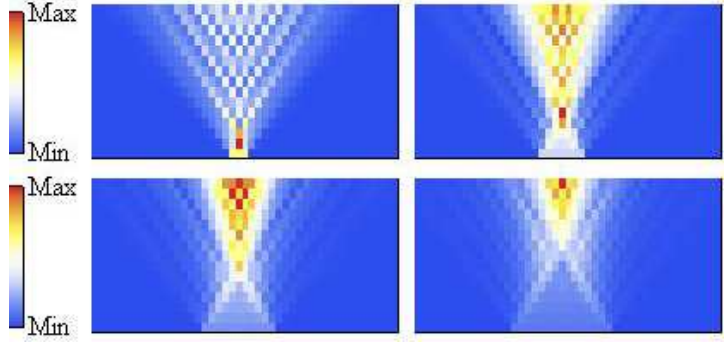} \end{center}
\caption{\label{fig:0}Space-time diagram of an evolving finite packet of cells. The panels correspond to $N=1, 3, 5, 7$  from top left to right bottom, and $\Delta=0.05$. Emergent focusing can be distinguished at $N=7$, where the width of the focusing region is smaller than $2N+1$.}
\end{figure}

\begin{figure}[h!]
\begin{center}  \includegraphics[width=9.5cm]{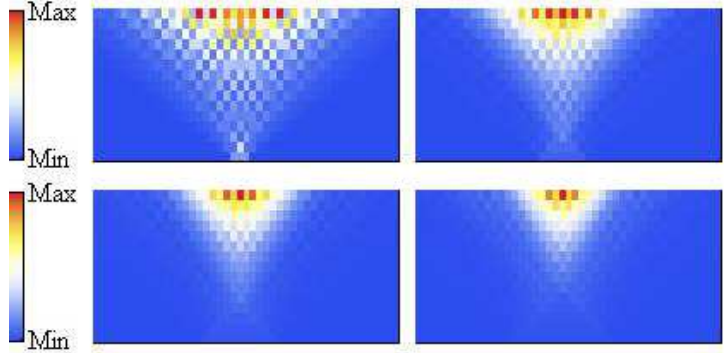} \end{center}
\caption{\label{fig:1}Space-time diagram of an evolving finite packet for $\Delta=0.07$. As before, the panels correspond to $N=1,3,5,7$, but no emergent focusing can be distinguished, as there is no contraction of the packet at a local maximum of $|\phi|^2$.}
\end{figure}

\bea
x_{\tau+1,n} = \Delta (y_{\tau,n+1} + y_{\tau,n-1}) + x_{\tau,n},
\label{0}
\eea
\bea
y_{\tau+1,n} = -\Delta (x_{\tau,n+1} + x_{\tau,n-1}) + y_{\tau,n}, 
\label{1}
\eea
where $\Delta$ is a real (coupling) constant. The solutions are obtained most easily by a change of variables $\phi_{\tau, n} = x_{\tau,n} + i y_{\tau,n}$, leading to a single complex equation

\bea
i (\phi_{\tau+1,n} - \phi_{\tau,n} ) = \Delta(\phi_{\tau,n+1} + \phi_{\tau,n-1} ).
\label{2}
\eea
Evidently, $\phi_{\tau+1,n}$ can be put in terms of itself at a previous time step using space translation operators $T_n, T_n^{\dagger}$. This shows that (\ref{2}) can be expressed as a Floquet operator:

\bea
\phi_{\tau+1,n} = U_{\scriptsize \mbox{discrete}}  \cdot \phi_{\tau, n}, 
\label{2.1}
\eea
\bea
U_{\scriptsize \mbox{discrete}} = \v 1 - i \Delta (T_n + T^{\dagger}_n).
\label{3}
\eea
It must be stressed, however, that $U_{\scriptsize \mbox{discrete}}$ is not unitary. In order to solve any initial data problem and find diffractive behaviour, we must express $\psi_{\tau,n}$ as a linear superposition of independent solutions, showing explicitly the dependence of the expansion coefficients on $\psi_{0,n}$. A basis of separable solutions in $\tau$ and $n$ is given by

\bea
\psi^{k}_{\tau,n} = \ncal \mbox{e}^{ikn} (1-2i\Delta \cos k)^{\tau}.
\label{4}
\eea
Without bothering with the normalization $\ncal$, we can see that these solutions ressemble Bloch waves in space, but the temporal part is not unimodular, in agreement with (\ref{3}) -- except for the trivial case $\Delta=0$, which shall be studied later as a continuous limit.

The solutions can be written as

\bea
\phi_{\tau,n} = \frac{1}{2\pi} \int_{-\pi}^{\pi} dk \quad C_k \psi^{k}_{\tau,n}
\label{5}
\eea
with $C_k$ given in terms of initial data in the form

\bea
C_k = \sum_{m= -\infty}^{\infty} \phi_{0,m} \mbox{e}^{-ikm}.
\label{6}
\eea
By substitution of (\ref{6}) in (\ref{5}), we conclude that any solution can be reached by the application of a discrete kernel

\bea
\phi_{\tau,n} =  \sum_{m= -\infty}^{\infty} K_{\tau; n-m} \quad  \phi_{0,m}
\label{7}
\eea
where $K$ can be easily computed:
\bea
K_{\tau; n-m} = \frac{1}{2\pi} \int_{-\infty}^{\infty} dk \quad \mbox{e}^{ik(n-m)} (1-2i\Delta \cos k)^{\tau}  \nonumber \\
= \sum_{ \scriptsize \begin{array}{c} r=0 \\  r+n-m \quad \mbox{even}\end{array}}^{\tau} \frac{\tau! (-i\Delta)^r}{(\tau-r)! \left( \frac{r+n-m}{2} \right)! \left( \frac{r+m-n}{2} \right)!}.
\label{8}
\eea
This expression for $K$ is a polynomial in $\Delta$ and it is therefore a closed result. 

We are particularly interested in the propagation of a square distribution

\bea
\phi_{0,n} = \cases{1& if $n \leq N$ \\ 0&\textrm{otherwise}},
\label{8.1}
\eea
which can be established also in closed form. If the initial cloud has length $2N+1$ and is centered around the origin, the linear automaton is solved by

\bea
\phi_{\tau,n}= \frac{1}{\sqrt{2N+1}} \sum_{m=-N}^{N} \sum_{ \scriptsize \begin{array}{c} r=0 \\  r+n-m \quad \mbox{even}\end{array}}^{\tau}   \frac{\tau! (-i\Delta)^r}{(\tau-r)! \left( \frac{r+n-m}{2} \right)! \left( \frac{r+m-n}{2} \right)!}. 
\label{9}
\eea
The results are displayed in figures \ref{fig:0} and  \ref{fig:1} as functions of space and time. The intensity $x_{\tau,n}^2 + y_{\tau,n}^2$ has a transition from ever increasing solutions (non-unitary evolution) to a short-time focusing behaviour. For values of $\Delta < 0.05$, only wavepackets with $N \geq 7$ show a contraction and an expansion, i.e. focusing emerges as a function of the intial number of lit cells (fig. \ref{fig:0}). On the other hand, if $\Delta > 0.05$, the evolution produces an increasing intensity function everywhere, with no trace of a maximum at a minimal width (fig. \ref{fig:1}). We explain the anatomy of expanding and contracting distributions in fig. \ref{fig:1.2}.

\subsection{Emergent focusing in tight-binding models}

\begin{figure}[t]
\begin{center}  \includegraphics[width=15cm]{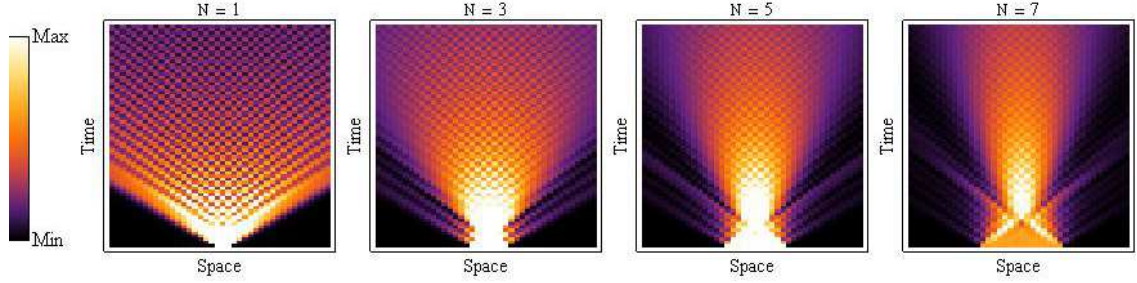} \end{center}
\caption{\label{fig:2}Intensity pattern in space-time for a square packet in a tight-binding model. Once again, we find that the emergence of focusing corresponds to $N \approx 7$ in the last panel.}
\end{figure}

\begin{figure}[t]
\begin{center}  \includegraphics[width=9.5cm]{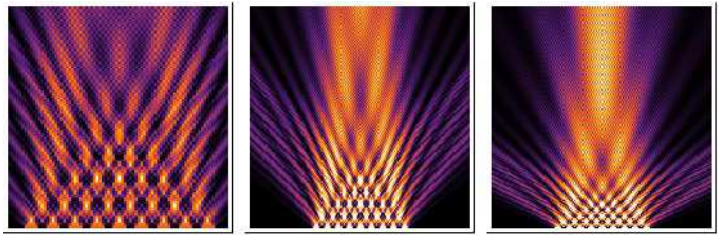} \end{center}
\caption{\label{fig:3}Intensity patterns resulting from an array of 9 square packets with $N=2$ each. The panels show the evolution at different scales, where the rightmost is an unequivocal depiction of refocusing.}
\end{figure}

The typical realization of a NNTBM comes in the form of single-band systems made of periodic arrays of potential wells (quantum-mechanical waves) or dielectric slabs (electromagnetic modes). In our previous model, we may write $\Delta = T \Lambda$ and divide both sides of (\ref{2}) by $T$. In the limit $T \rightarrow 0, \tau \rightarrow \infty$ with a fixed continuous time $t = T\tau$, we recover 

\bea
i \frac{\partial \phi_n(t)}{\partial t} = \Lambda \left[\phi_{n+1}(t)+ \phi_{n-1}(t) \right].
\label{10}
\eea
The physical meaning of $\phi_n$ comes from the expansion of any wavefunction $\phi$ in terms of a  (single band) Wannier basis $| n \>$ with crystal site number $n$

\bea
\phi(x) = \sum_{n} \phi_n \< x | n \>.
\label{10.1}
\eea 
The propagation problem posed by (\ref{10}) can be solved now by the well-known kernel \cite{sadurni2012(2)}, \cite{sadurni2013}
\bea
K_{n-m}(t) = i^{n-m} J_{n-m} (\Lambda t),
\label{11}
\eea
which satisfies unitarity. The propagation of wavepackets is reached by the formula

\bea
\phi_n(t) = \sum_{m=-\infty}^{\infty} i^{n-m} J_{n-m}(\Lambda t)\phi_m(0),
\label{12}
\eea
and for a square packet of length $2N+1$, i.e. 
\bea
\phi_n(0)= \frac{1}{\sqrt{2N+1}}\cases{
1 & if $n \leq N$ \\
0 & otherwise}
\label{12.1}
\eea
we have the finite sum

\bea
\phi_{n}(t) = \frac{1}{\sqrt{2N+1}}\sum_{m=-N}^{N} i^{n-m} J_{n-m}(\Lambda t).
\label{13}
\eea
The results of the evolution are displayed in figure \ref{fig:2} and in units $\Lambda=1$, which shows that the entire phenomenon depends exclusively on the number $N$. Once again, we find that for $N \geq 7$ the features of diffractive focusing are well developed: we find a regime of contraction accompanied by oscillations, a bright zone of maximum intensity at a finite time and a regime of parabolic expansion for long times (violet fringes) instead of the conical caustics that are typical of pulses in discrete systems \cite{sadurni2013}. We shall see that $N=7$ in fact corresponds to a critical size for which a {\it measure\ }saturates (fig. \ref{fig:5.1}). 

We may further explore the formation of Talbot carpets \cite{case2009}, \cite{kim2010}, \cite{turlapov2005}, but we do so in discrete space and using our propagator \cite{sadurni2012(2)}. In the case of electromagnetic waves one can achieve periodic boundary conditions by means of mirrors judiciously placed  next to a single slit. However, in quantum mechanical realizations such as matterwaves in optical traps, it is necessary to prepare many squarepackets with the aim of producing locally periodic evolution at the center of the array. But it turns out that the finite size of such an experiment must lead again to a refocusing of the total distribution, as depicted in figure \ref{fig:3}. Therefore, we may regard the time previous to diffractive focusing as a {\it mini-carpet.\ }  

It is worth mentioning that Talbot carpets in continuous variables have been identified also as an emergent phenomenon \cite{berry2001}. The type of emergence that is alluded to in such reference is related to fractality in the limit of many slits, rather than in a continuous limit. 

\subsection{Focusing in the continuous limit}

The 1DSE exhibits diffraction in time \cite{moshinsky1952} for sharp edges, and two of such discontinuities produce focusing {\it independently of the scale,\it } i.e. the packet concentrates at a fixed value of a dimensionless variable which is proportional to the square of the size. This well-known result \cite{case2012}, \cite{sadurni2012} can be obtained also as a limit of the previous NNTBM by a careful definition of limits. Here it is convenient to recall that the conical caustics emerging from pulses in NNTBM are related to effective relativistic equations. These effective theories have been identified and extended to hexagonal lattices \cite{sadurni2010}. How do we recover a non-relativistic Schr\"odinger equation? We take the limits in the following way:

\bea
|\Lambda| \rightarrow \infty, \quad
|n| \rightarrow \infty, \quad
\frac{n}{\sqrt{|\Lambda}|} \rightarrow x \in \Re.
\label{14}
\eea
If $\Lambda <0$, the resulting equation is 

\bea
 i\frac{\partial \phi(x,t)}{\partial t} = -\nabla^2 \phi(x,t),
\label{14.1}
\eea
which is a dimensionless Schr\"odinger equation. This limit also reduces the discrete propagator to the usual gaussian kernel, as can be proved using the Meissel expansion for Bessel functions \cite{watson1922}.

It is important to recall that the focusing time of the square packet is given by 

\bea
t \approx 0.026 \times L^2
\label{14.2}
\eea
where $L$ is the size of the distribution. This implies that for any value of $L$ we can find a regime in which the maximum occurs.

\section{Quantification of the transition}

In this section we define a width in terms of probabilities, only for wavepackets in NNTBM.

\subsection{Remarks on criticality}

Wavepacket concentration or segregation can be described in many ways, but we should not content ourselves with a mere visualization of the phenomenon in space and time. In a previous work \cite{case2012} we dealt with a new measure capable of describing the width of a distribution. Such a gaussian filter was shown to decrease at a critical time, in opposition to the second moment inherent to quantum-mechanical probability distributions. In this occasion, we propose a simpler, physically intuitive quantity to gauge the concentration of a packet: the probability of permanence in a small box as a function of its width. This simple filter is nothing but the expectation value of a Heaviside function, and it will capture the contraction of quantum-mechanical distributions if it increases as a function of time for certain values of the box width and the wavepacket size. Incidentally, this measure is also intuitive for electromagnetic waves, since it corresponds to the field energy stored in the box. Criticality can be identified in this way as a transition in which a maximum in the probability of permanence occurs at a finite time. In the absence of such a maximum, we can fairly say that diffraction phenomena do not take place, and that the system is far from being described by a continuous wave equation. From this perspective, we may also note that wave dynamics {\it is\ } an emergent phenomenon in which the appearence of travelling signals or pulses is a necessary condition, but not a sufficient one. The picture is obviously completed by diffraction. 

In connection with phase transitions \cite{lebowitz1999}, we underscore the fact that diffraction is a time- and space-dependent phenomenon, rather than a process described by a phase diagram near equilibria. Strictly speaking, we should not look for a phase transition in the traditional sense (discontinuities in thermodynamic variables), but we may still use a dynamical quantity to characterize a transient aggregation. Moreover, the elimination of the time variable can be achieved in our study by asking whether an increase in the probability of permanence happens or not, depending only on the size of the wavepacket. 

\begin{figure}[t]
\begin{center}  \includegraphics[width=8.5cm]{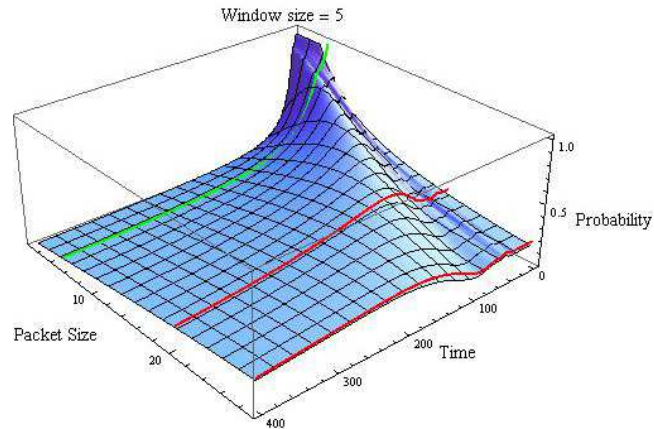} \end{center}
\caption{\label{fig:4} The function $1-\wcal_{\scriptsize \mbox{discrete}}$ for $N_0 = 5$. The green curve always decreases with time, whereas the red curves illustrate focusing by an increase at early times until a maximum is reached.}
\end{figure}
\begin{figure}[t]
\begin{center} \includegraphics[width=8.5cm]{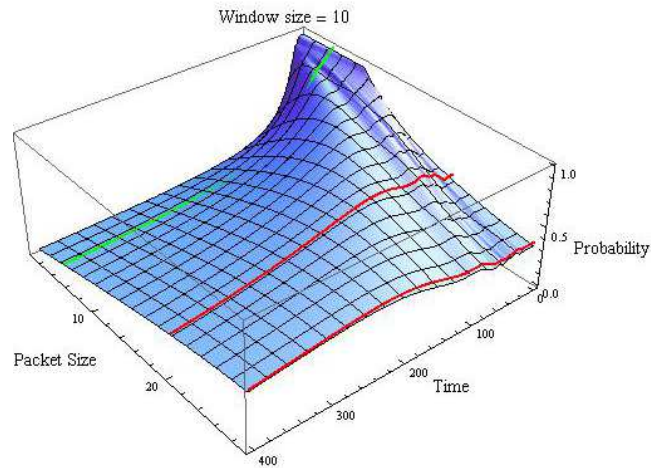} \end{center}
\caption{\label{fig:5} The function $1-\wcal_{\scriptsize \mbox{discrete}}$ for $N_0 = 10$. As before, the regimes indicated by green and red curves can be distinguished.}
\end{figure}

\begin{figure}[h!]
\begin{center} \includegraphics[width=8.5cm]{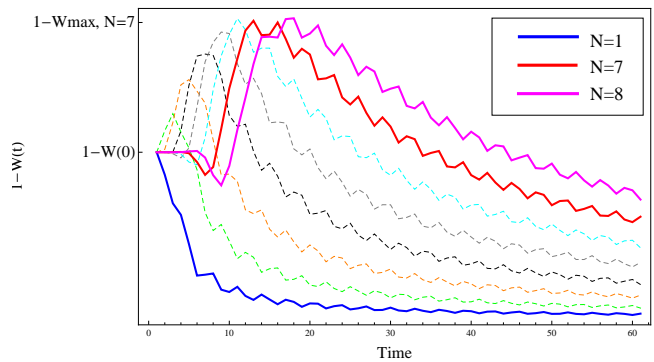} \end{center}
\caption{\label{fig:5.1} The function $\left[1-\wcal_{\scriptsize \mbox{discrete}}(t)\right]/\left[1-\wcal_{\scriptsize \mbox{discrete}}(0)\right]$ 
for minimum window size $N_0 = 1$ and packet sizes $N=1,...,8$. The transition occurs at $N=7$ (red curve), since for all $N \geq 8$ the maximum of the curve has the same value. For $N=1$ (blue curve) there is no focusing.}
\end{figure}

\subsection{A definition of width $\wcal$}

For our wavefunction in (\ref{13}), we define our measure of width $\wcal$ such that

\bea
1-\wcal_{\scriptsize \mbox{discrete}}(N_0, N; t) = \sum_{n=-N_0}^{N_0} |\psi_{n}(t)|^2.
\label{15}
\eea
When $N_0 < N$, expression (\ref{15}) corresponds to the probability that a quantum-mechanical particle enters into a window smaller than the initial width $2N+1$. In the electromagnetic case, the r.h.s. of (\ref{15}) gives the energy of the field stored in an interval smaller than the slit width. The results in figs. \ref{4} and \ref{5} show that a transition indeed happens: when the wavepacket reaches a certain size ($N \approx 7$) the probability of permanence increases as a function of time (red curves) until it reaches a maximum (focusing) and then decreases monotonically (expansion). For smaller packets, there is a regime in which the probability always decreases (green curve). The existence of a transition is independent of the window size $2N_0 +1$; we show the results for the values $N_0 = 5, 10$ by way of example. 
It is important to mention that in the continuous case -- replacing sums by integrals in (\ref{15}) -- the measure would also show an apparent transition as a function of $N$ relative to $N_0$, but one should recognize that there is always focusing in this case, therefore there always exists a window size $N_0$ for a given $N$ such that (\ref{15}) increases. This feature is not shared by the discrete example, since for very small packets we do not see focusing; the measure shows it by displaying a decreasing behaviour even for the smallest possible value of $N_0$, if the wavepackets are small -- see figure \ref{fig:5.1}. On the other hand, for wavepackets $N \geq 7$, the function $\left[1-\wcal_{\scriptsize \mbox{discrete}}(t)\right]/\left[1-\wcal_{\scriptsize \mbox{discrete}}(0)\right]$ reaches a maximum whose value is the same for all sizes: this defines a transition.

\section{Wigner functions of discrete and continuous variables}

One of the challenges posed by wavepacket dynamics \cite{andreata2003}, \cite{bialynickibirula2002} is to predict a possible contraction from full knowledge of the initial condition. To this effect, a number of works have been devoted to the analysis of Wigner functions \cite{wigner1932}, \cite{case2008}. Phase space in quantum mechanics \cite{schleich2001}, \cite{olivares2012} has been useful in providing a pictorial description of evolution using space and momentum quasi-distributions; after all, complex wave functions contain information about probability and current distributions indicating probability flow. 

\begin{figure}[b]
\begin{center}  \includegraphics[width=11cm]{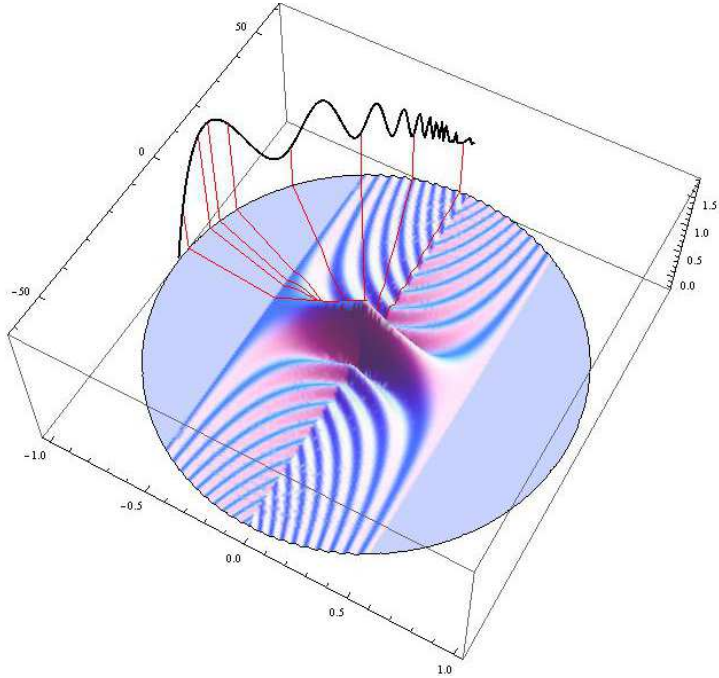} \end{center}
\caption{\label{fig:6} Wigner function in the $x,p$ plane and the intensity of the wavefunction at the origin (black curve) as the result of integration along the evolved $x,p$ variables (red lines). A global maximum can be noted.}
\end{figure}

A continuous Wigner function for a square packet distribution has been previously obtained and analyzed \cite{schleich2014}. The active point of view of phase space evolution indicates that the initial Wigner function must remain fixed in time, while the space undergoes a shear transformation $x(t)= -t p_0 + x_0, p(t) = p_0$. Then, by integrating over the line $x(t) = -t p_0$, one finds the emergence of a maximum for a critical time $t=t_c$, corresponding to the focusing time. We show the results in figure \ref{fig:6}.

We would like to find similar results for Wigner functions in a tight-binding array, with the purpose of making a just comparison and characterize the dynamics.

\subsection{Tight-binding Wigner functions and periodic phase space}

A straightforward generalization of Wigner functions for tight-binding arrays can be obtained by taking the following expression \cite{case2008} as a starting point ($\hbar = 1$):

\bea
W(x,p)= \int_{-\infty}^{\infty} dx' \mbox{e}^{ipx'} \phi(x-x'/2,t) \phi^{*}(x+x'/2,t) \nonumber \\
= 2 \int_{-\infty}^{\infty} d\xi \mbox{e}^{2ip\xi} \phi(x-\xi,t) \phi^{*}(x+\xi,t).  
\label{16}
\eea
With this innocent change of integration variable, we can define now

\bea
W_n(k; t)= 2 \sum_{m=-\infty}^{\infty} \mbox{e}^{2ikm} \phi_{n-m}(t) \phi^{*}_{n+m}(t),
\label{17}
\eea
where $k$ is now Bloch's quasi-momentum. This function has a number of properties that are unique to discrete space. For example, $W_n(k;t)$ is periodic in $k$ with a repetition of the strip $-\pi/2 \leq k \leq \pi/2$ (this is the famous Wigner-Seitz cell, another concept introduced by Wigner). For a Bloch wave of quasi-momentum $q$ at $t=0$, $W_n(k;0)$ is given by periodically spaced Dirac deltas centered at $q + 2\pi m$, whereas for a point-like function $\phi$ one finds that $W_n(k;0)$ is given by a Kronecker delta. Another way to express (\ref{17}) comes from Bloch expansions of the form

\bea
\phi_n(t) = \int_{-\pi}^{\pi} dk' \mbox{e}^{ik'n} \tilde \phi(k',t),
\label{18}
\eea
which lead, upon substitution in (\ref{17}) and a two-dimensional change of variables in $k',k''$, to the following representation

\bea
W_{n}(k;t) = \int_{-\pi}^{\pi} dk' \mbox{e}^{2ik' m} \tilde \phi(k+k',t) \tilde \phi^{*}(k-k',t).
\label{19}
\eea
It also happens that if $\phi$ is real, $W_n(k;0)$ is an even function of $k$. Moreover, the evolution of real wavepackets gives rise to complex functions in general, and we should expect a broken symmetry under $k \mapsto -k$ due to evolution.
 
\begin{figure}[t]
\begin{center}  \includegraphics[width=14cm]{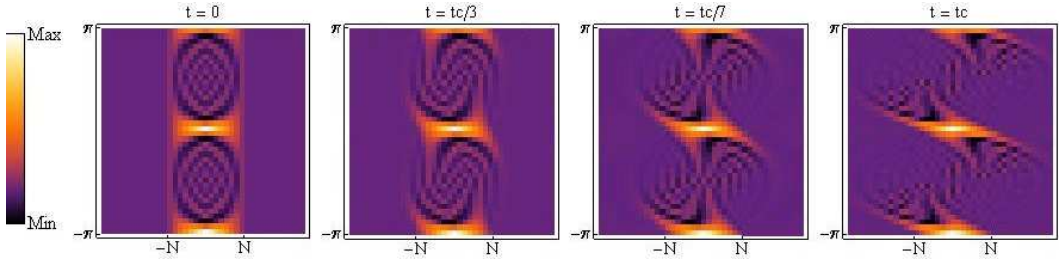} \end{center}
\caption{\label{fig:7} Evolution of the Wigner function in the $n,k$ plane, for $N=7$. The evolution ressembles a shear near the origin, but the shape is considerably modified around $k=\pm \pi/2$ due to periodicity. The focusing is reached in the last panel.}
\end{figure}

\begin{figure}[t]
\begin{center} \includegraphics[width=14cm]{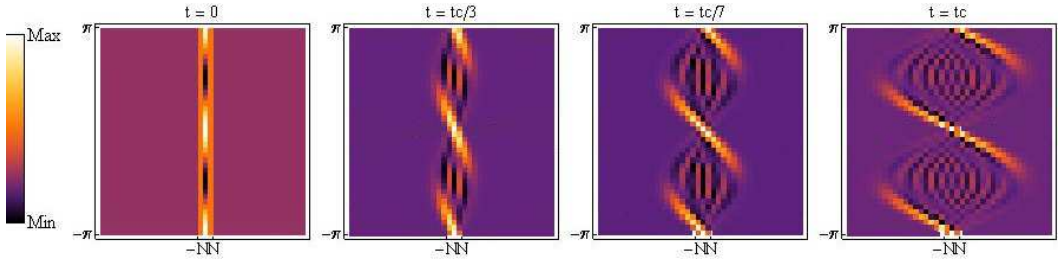} \end{center}
\caption{\label{fig:8} Evolution of a tight-binding Wigner function for a small wavepacket ($N=1$). Although there is a shear transformation near the origin, alternating signs remain along the line $n=0$ for all times, giving rise to destructive interference and no focusing.}
\end{figure}

\begin{figure}[t]
\begin{center}  \includegraphics[width=14cm]{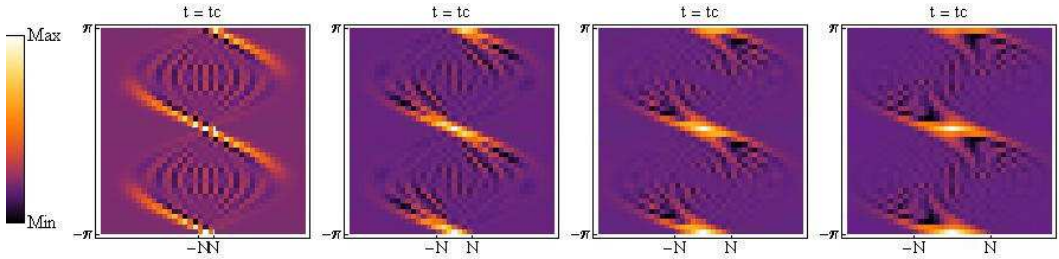} \end{center}
\caption{\label{fig:9} Wigner function in the $n,k$ plane evaluated at focusing time for various sizes. Focusing is reached at $N=7$, corresponding to the rightmost panel.}
\end{figure}

Since we have the time-dependent wavefunction (\ref{13}) at our disposal, it is easy to evaluate (\ref{17}) numerically and draw some conclusions in connection with focusing. At $t=0$ we have

\bea
W_n(k;0) = \Theta(N-|n|) \frac{\sin(k \left[ 2N-2|n|+1\right])}{\sin k}.
\label{20}
\eea
At later times, it seems that large wavelengths and small values of $k$ should help us to recover the evolution of the usual Wigner function in continuous space. In figures \ref{fig:7}, \ref{fig:8} and \ref{fig:9} we show the corresponding time development, which ressembles a shear transformation for $k << \pi/2$. However, if $k\sim \pi/2$, the periodicity of the Wigner function produces a completely different map. By comparing figs. \ref{fig:7} and \ref{fig:8}, we observe that around such a point the patterns suffer a serious deformation for $N=7$ (critical width), but they undergo a mere expansion for $N=1$. The transition from small $N$ to large $N$ is given in fig. \ref{fig:9}, where the emergence of focusing can be identified with a gradual vanishing of $W_n(k;t)$ in the region of alternating signs around $k \sim \pm \pi/2$, as well as a growth of the region around $(n,k)=(0,0)$ in which $W_n(k;t)$ is strictly positive.

We complete our analysis by finding an approximate expression for the evolution of $W_n(k;t)$ in terms of $W_n(k;0)$.

\begin{figure}[h!]
\begin{center}  \includegraphics[width=8cm]{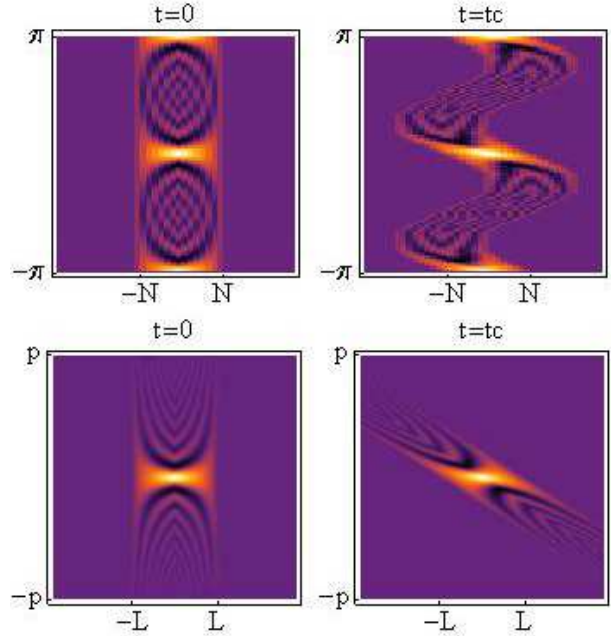} \end{center}
\caption{\label{fig:10}Upper row: Tight-binding Wigner function for $N=7$ and its evolution at focusing time using aproximation (\ref{27}). Lower row: Continuous Wigner function and its evolution at focusing time. The zones around $(n,k)=(0,0)$ and $(x,p)=(0,0)$ are similar, explaining the focusing effect for the discrete case.}
\end{figure} 

\subsubsection{Equations of motion for tight-binding Wigner functions}

The shear transformation operating on continuous variables $x,p$ can be used to prove that $W(x,p;t)$ satisfies a first order differential equation. With the choice $2m=-1$ in the dimensionful Schr\"odinger equation -- which corresponds to the continuous limit of a tight-binding model with {\it positive\ }nearest-neighbour couplings -- we have

\bea
\frac{\partial W(x,p;t)}{\partial x} -\frac{1}{2p}\frac{\partial W(x,p;t)}{\partial t} = 0,
\label{21}
\eea
and this leads immediately to solutions of the form $W(x,p;t)= W(x + 2p t, p;0)$. 

In a tight-binding chain, the evolution equation of $W_n(k;t)$ must be a finite difference equation in space and a differential equation in time. We may start our derivation by recalling that $\phi_n(t)$ satisfies (\ref{10}), compelling us to compute the time derivative $ \partial W_n(k;t) / \partial t$. Unfortunately, we do not obtain a closed result in this first step, and (\ref{21}) can no longer be valid. Computing a second derivative, however, leads to a very appealing result

\bea
\frac{1}{4 \Lambda^2 \sin^2 k} \frac{\partial^2 W_n(k;t)}{\partial t^2} - W_{n+1}(k;t) - W_{n-1}(k;t) + 2W_n(k;t) = 0. \nonumber \\
\label{22}
\eea 
This equation can be proved by using (\ref{10}) and a change of summation indices in (\ref{17}). Equation (\ref{22}) is closely related to a wave equation, if we recognize that the combination of translation operators $T_n + T_n^{\dagger}- 2 \v I$ corresponds to the discretization of a second derivative in space. We have

\bea
\left[\frac{1}{(2 \Lambda \sin k)^2} \frac{\partial^2}{\partial t^2} - \left( T_n + T_n^{\dagger}- 2 \v I \right)\right]W_n(k;t) = 0, 
\label{23}
\eea
with an effective wave velocity given by $2 |\Lambda \sin k|$. The solutions of (\ref{23}) are formally given as expansions in Bessel functions of an even index, i.e. if $W_n(k;t)$ is decomposed into positive and negative 'velocities' 

\bea
W_n(k;t) = W^{+}_n(k;t) + W^{-}_n(k;t)
\label{24}
\eea
we will have 
\bea
W^{\pm}_n(k;t) = \sum_{m=-\infty}^{\infty} W^{\pm}_m J_{2n+m} (\pm 2 \Lambda t \sin k ).
\label{25}
\eea
The expansion coefficients $ W^{\pm}_m$ are obviously fixed by the intial conditions $W_n(k;0)$ and $\partial W_n(k;0) / \partial t$, which are both known. Although the expansion (\ref{25}) is entirely correct and ready for a numerical evaluation, it is not known in closed form for our evolving square packet. An interesting approximation emerges in the strong coupling regime $|\Lambda| >> 1$, allowing us to set $\nu = n /|\Lambda|$ as a continuous variable, and with the more amenable notation $ W(\nu,k;t) = W_{|\Lambda| \nu}(k;t)$, equation (\ref{23}) becomes 

\bea
\frac{1}{(2 \sin k)^2} \frac{\partial^2 W(\nu,k;t) }{\partial t^2} -  \frac{\partial^2 W(\nu,k;t) }{\partial \nu^2} = 0. 
\label{26}
\eea
Now we may attempt a solution by considering only positive velocities, eliminating $W^{-}$ in (\ref{24}). We finally obtain

\bea
W(\nu, k; t) = W(\nu + 2 t \sin k, k; 0)
\label{27}
\eea
which is similar to a shear transformation, but applied to variables $\nu, \sin k$. It is also clear that if $\sin k \sim k$, we will recover the continuous result (\ref{21}). The evaluation of the function (\ref{27}) with $W_n(k;0)$ given by (\ref{20}) generates patterns that are quite similar to the numerical evaluation shown in figs. \ref{fig:7} and \ref{fig:8}. On the other hand, the approximation is inadequate for $k \sim \pi/2$, where the periodicity of $W_n(k;t)$ in $k$ must be imposed. A comparison of patterns is given in figure \ref{fig:10}.

\section{Discussion}

Our efforts have led us to characterize diffraction -- and in particular diffractive focusing -- as a phenomenon that emerges in the transition from discrete models to continuous wave equations. The collection of results that has been presented here demonstrates that complex dynamics may occur not only because of the nature of equations, but also because of special choices of intitial conditions. In the context of cellular automata, it is clear that an increasing number of participants gives rise to more intricate evolution, but we have shown that such a number is also responsible for emergent diffraction, establishing that complexity and diffractive focusing are correlated. Although we have restricted our attention to a specific measure in terms of plain probabilities, the existence of a transition as a function of $N$ should be present in any other convex function that acts as a filter. 

Experiments with coupled waveguides, dielectric slabs and matter waves in optical lattices can be proposed, with the hope that their future realization sheds more light on the effects produced by selective measurements.   

\ack

I would like to express my gratitude to Prof. W. P. Schleich and Prof. W. Case for useful discussions. Financial support from CONACyT under project CB2012-180585 is acknowledged.

\section*{References}

\end{document}